\documentclass[aps,pre,twocolumn,superscriptaddress]{revtex4-2}
\usepackage[T1,T2A]{fontenc}
\usepackage[utf8]{inputenc}
\usepackage[russian,english]{babel}
\usepackage{graphicx}
\usepackage{dcolumn}
\usepackage{bm}
\usepackage{graphicx}
\usepackage{amsmath}
\usepackage{amssymb}
\usepackage{pifont}
\usepackage{epstopdf}
\usepackage{color}

\newcommand{\Da}{\textrm{Da}}

\begin{document}

\title{Diffusiophoresis of 
 ionic catalytic particles}
\author{Evgeny S. Asmolov }
\email[Corresponding author: ]{aes50@yandex.ru}
\affiliation{Frumkin Institute of Physical Chemistry and Electrochemistry, Russian Academy of Sciences, 31 Leninsky Prospect, 119071 Moscow, Russia}

\author{Olga I. Vinogradova}
\affiliation{Frumkin Institute of Physical Chemistry and Electrochemistry, Russian Academy of Sciences, 31 Leninsky Prospect, 119071 Moscow, Russia}

\begin{abstract}
A  migration of charged particles relative to a solvent, caused by a gradient of salt concentration and termed a diffusiophoresis, is of much interest being exploited in many fields. Existing theories  deal with diffusiophoresis of passive inert  particles.
In this paper, we extend prior models by focusing on a particle, which is both passive and catalytic, by postulating an uniform  ion release over its surface.
We derive an  expression for a particle velocity depending on a dimensionless ion flux (Damk\"{o}hler number Da) and show that a charged region is formed at distances of the order of the particle size, provided the diffusion coefficients of anions and cations are unequal. When Da becomes large enough, the contribution of this (outer) region to the particle velocity dominates. In this case the speed of catalytic passive particles augments linearly with Da and is inversely proportional to the square of electrolyte concentration. As a result, they  always migrate towards a high concentration region and in dilute solutions become much faster  than inert (non-catalytic) ones.
\end{abstract}

\maketitle


\section{Introduction}

The migration of colloidal particles in a molecular~\cite{anderson1982motion} or an ionic~\cite{deryagin1961,prieve1984motion} solution,
caused by a gradient in concentration of solutes and termed a diffusiophoresis, is currently a subject of active research.
Early work has  assumed that particles are chemically uniform, inert (i.e. no chemical reactions at the surface accompanied
by an ion release occur) and, therefore, passive. The diffusiophoresis of such passive particles is currently widely employed
for their manipulation~\cite{abecassis2009osmotic,velegol2016,shim2022diffusiophoresis}. However, it was later understood that  a
diffusiophoretic migration can also be induced by particles themselves~\cite%
{paxton2005motility,moran2010locomotion}. In this case, termed a
self-diffusiophoresis, the concentration gradients are generated by an
active catalytic particle that nonuniformly releases ions from its surface.
A phenomenon of self-diffusiophoresis is an origin of the propulsion of
Janus swimmers~\cite{golestanian2007,moran2017,asmolov2022COCIS}. Such
swimmers already find numerous applications, for example, drug delivery,
lab-on-a-chip devices, nano-robotics, and more~\cite{li2017micro,hu2020micro}.

There is a large literature describing attempts to provide a satisfactory theory of the diffusiophoretic migration
 in electrolyte solutions. We mention below what we believe are the more relevant contributions. It is currently well recognized
 that in the case of  such a ionic
diffusiophoresis, the concentration gradients of cations and anions induce
an (external) electric field to hold an electroneutrality (zero current) of
a solution, which, in turn, slows down faster ions and accelerates slower
ones. Thus, the diffusiophoresis of charged particles in electrolyte
solution is a consequence of a combined effect of concentration gradients
and electric field~\cite{deryagin1961,prieve1984motion}. It is also well understood that the electrostatic diffuse layer (EDL), i.e., the region
where the surface charge is balanced by the cloud of
counterions, is an origin of an ionic diffusiophoresis. So far, most attention has been focussed on the case of thin (compared to the particle radius $a$) EDL that is well justified for microparticles. In this situation, it appears that the liquid slips over the particle surface. The emergence of such an apparent slip flow has been used to obtain a propulsion velocity without tedious calculations since in this case the system of governing equations (for ion concentrations,
electric potential, and fluid velocity) can be solved  with the method of matched asymptotic expansions.  Some solutions are already known for
passive inert~\cite{prieve1984motion,anderson1989colloid,pawar1993} and
active catalytic~\cite{yariv2011electrokinetic,sabass2012nonlinear,
nourhani2015,ibrahim2017,de2020self,asmolov2022self} particles. Note that although the propulsion mechanism appears similar for both types of
particles, there is some important difference. The point is that catalytic swimmers generate a supplementary concentration gradient outside the EDL, and this (weakly charged) outer region extends to distances of the order of $a$~\cite{nourhani2015,asmolov2022self} by controlling a flow within EDL (i.e. in the inner
region). Finally, we mention that there exist two alternative approaches to modeling chemical activity at the
particle surface. One represents the so-called
kinetic-based model, where ion fluxes depend on the instant local
concentrations, and requires a detailed information on the rates of surface
chemical reactions~\cite{moran2011electrokinetic,yariv2011electrokinetic,sabass2012nonlinear,ibrahim2017}. Another approach is based on the flux-based model, which will be employed here, where the surface ion fluxes
are prescribed \cite{nourhani2015,de2020self,asmolov2022self}. The advantage
of the latter approach is that it does simplify the analysis of
steady-state processes.

Yet, despite those advances, at least one experimentally relevant system has not received an attention. A variety of particles can generate
a uniform ionic flux from their surface. This is expected when small grains of salt dissolve in water~\cite{mcdermott2012}, has been observed for   porous microparticles in illuminated with an appropriate
wavelength solutions of a photosensitive ionic surfactant~\cite{feldmann.d:2020}, or could be caused by a
catalytic reaction at the uniform surface~\cite{dey2015micromotors,ma2016enzyme,patino2018fundamental}. For instance, it has been found the
uniformly coated by urease polymer microspheres decompose urea and migrate towards its higher  concentration
\cite{dey2015micromotors}. To the best of our knowledge the diffusiophoretic migration of such particles has never been treated theoretically. It
is of considerable interest to explore what happens with a passive catalytic particle
placed to a solution with a salt concentration gradient, since it should
induce the flow not only inside the EDL, but also in the outer region, but
we are unaware of any previous work that has addressed this question.

In the present paper we consider a diffusiophoresis of catalytic particles that remain passive, since an ion release is uniform.
We develop a theory that describes the diffusiophoretic velocity of such
particles. Our results show that their migration velocity include contributions associated with the inner and outer regions. The
former dominates at the low Damk\"{o}hler numbers leading to a migration
controlled by the surface potential as for inert particles. The latter prevails at the high Damk%
\"{o}hler numbers. In this case the diffusiophoretic migration does not
depend on the surface potential and all particles propel fast towards a
region of higher salt concentration.

Our paper is organized as follows. Section~\ref{s2} describes the theory. We
present our model and governing equations, describe their solution in the
outer region, and give a detailed derivation of equations for the
diffusiophoretic velocity. Section~\ref{s3} contains the results of
numerical calculations, validating the theoretical predictions. We conclude
in Sec.~\ref{s4}.

\section{Theory}

\label{s2}

\begin{figure}[h]
\begin{minipage}[h]{1.\linewidth}
\includegraphics[width=0.9\columnwidth ]{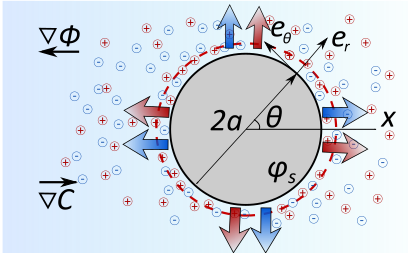} 
\end{minipage}
\vfill
\caption{Sketch of diffusiophoresis of catalytic particle and coordinate
system. Dashed circle shows a border of a thin EDL.}
\label{fig:sketch}
\end{figure}

Our aim is to calculate the velocity $\mathbf{u}$ of the steady-state migration of a catalytic spherical particle of radius $a$ immersed in an 1:1 aqueous
electrolyte solution of temperature $T$ and permittivity $\epsilon$. The bulk solution is characterized by a (small) concentration gradient in the $x-$direction (see Fig. \ref{fig:sketch}).
 A number density (concentration) of ions in the bulk is $c_{\infty }\left(
1+\varepsilon x\right)$, where the origin of coordinates is defined at the center of the particle,
$\varepsilon =\dfrac{1}{c_{\infty }}\dfrac{%
dc_{\infty }}{dx}\ll 1$, and coordinates are scaled by $a.$
Fluxes of emitted
cations and anions are assumed to be equal and uniform along the particle surface. %
Thus  finite gradients of concentration and electric potential are induced, but being  spherically symmetric they cannot cause the  particle motion. However, a combination of large symmetric and small external gradients in the $x$-direction generates a propulsion with the velocity that scales with $\varepsilon .$ So, the
mathematical formulation is similar to that for a non-catalytic particle, but the symmetric gradients are different. 

The reference Debye screening length of a bulk solution, $\lambda_D=\left( 8\pi \ell _{B}c_{\infty}\right) ^{-1/2}$, is defined at the plane $x=0$ using the Bjerrum
length, $\ell _{B}=\dfrac{e^{2}}{\epsilon k_{B}T}$, where $e$ is the elementary positive charge, and $k_{B}$ is the Boltzmann
constant. Upon increasing molar concentration from $10^{-6}$ to $10^{-1}$ mol/l the Debye length is reduced from about 300 down to 1 nm. Since the EDL extends to distances of the order of the Debye length, it is convenient to introduce the parameter $\lambda =\lambda _{D}/a$, which characterizes its relative thickness. Here we address the case of a thin EDL, i.e. require $\lambda \ll 1,$  which would
be realistic for concentrated solutions and/or microparticles.

The concentration gradient generates a diffusio-osmotic flow near the surface of a local velocity $\mathbf{v}$, which in turn provides
stresses that propel the particle. It is convenient to define a dimensionless fluid velocity $\mathbf{V}=\dfrac{4\pi \eta \ell _{B}a}{k_{B}T}\mathbf{v}$~\cite{saville1977}, where $\eta $ is the dynamic viscosity. For microparticles, the Reynolds and Peclet numbers are always small, so the Nernst-Planck and Navier-Stokes equations are decoupled
since the convective terms in them can be neglected.

 The dimensionless ion fluxes $\mathbf{J}^{\pm }$ are then governed by the
Nernst-Planck equations,
\begin{equation}
\nabla \cdot \mathbf{J}^{\pm }=0,  \label{NPH}
\end{equation}%
\begin{equation}
\mathbf{J}^{\pm }=-\nabla C^{\pm }\mp C^{\pm }\nabla \Phi,  \label{NP1}
\end{equation}%
with the local dimensionless concentrations $C^{\pm }$ scaled by $c_{\infty }$ and the electric potential $\Phi $ scaled by $k_{B}T/e \simeq 25$ mV.

The fluid flow satisfies the Stokes equations,
\begin{equation}
\mathbf{\nabla \cdot V}=0\mathbf{,\quad }\Delta \mathbf{V}-\mathbf{\nabla }P=%
\mathbf{f}.  \label{NS}
\end{equation}%
Here $P$ is the dimensionless pressure
(scaled by $\dfrac{k_{B}T}{4\pi \ell _{B}a^{2}}$) and $%
\mathbf{f}=-\Delta \Phi \mathbf{\nabla }\Phi $ is the electrostatic body
force.

Unequal distributions of oppositely charged ions generate an electric field
that satisfies the Poisson equation for the dimensionless electric potential,%
\begin{equation}
\Delta \Phi =-\lambda ^{-2}\frac{C^{+}-C^{-}}{2}.  \label{pois1}
\end{equation}%
Since $\lambda \ll 1,$ we construct the asymptotic solution using the method
of matched expansions in two regions with different lengthscales. The
lengthscale of the outer region is the particle size $a$ and that for the
inner region is the Debye length $\lambda _{D}$.

In the spherical coordinate system the boundary conditions for
concentrations at the particle surface ($r=1$) set the ion fluxes:%
\begin{equation}
\partial _{r}C^{\pm }\pm C^{\pm }\partial _{r}\Phi =-\mathrm{Da}\frac{D}{%
D^{\pm }},  \label{bc_1}
\end{equation}%
where
\begin{equation}
\mathrm{Da}=\frac{j a}{D c_{\infty }},\quad D=\frac{2D^{+}D^{-}}{D^{+}+D^{-}}.
\label{Da_def}
\end{equation}%
Here $\mathrm{Da}$ is the so-called Damk\"{o}hler number, which depends on   a dimensional flux
of released ions $j$ and their diffusion coefficients $D^{\pm }$. More precisely, the  Damk\"{o}hler number represents the ratio of the surface reaction rate to
the diffusive transfer rate~\cite{cordova2008,moran2017}.
It would be worthwhile to emphasize that such a definition of $\mathrm{Da}$ implies that it is small at high salt and/or when the surface flux is
small. By contrast, low $c_{\infty }$, especially along with a
large surface flux, would provide a large value of $\mathrm{Da}$.

Equation~\eqref{bc_1} can be reformulated as
\begin{equation}
\partial _{r}C^{\pm }\pm C^{\pm }\partial _{r}\Phi =-\mathrm{Da}\frac{1\mp \beta }{2},  \label{bc_11}
\end{equation}%
where we have introduced the diffusivity difference factor \cite{prieve1984motion,velegol2016}%
\begin{equation}
\beta =\frac{D^{+}-D^{-}}{D^{+}+D^{-}},  \label{bet}
\end{equation}%
which can vary from -1 to 1. For instance, in the case of salt grain dissolution, $\beta \simeq -0.207$ for NaCl and is much smaller, $\beta
\simeq -0.019,$ for KCl. The products of the enzymatic decomposition of
urea at the surface of functionalized particles form ions NH$_{4}^{+}$%
 and OH$^{-}$ \cite{de2020self} which yield $\beta \simeq -0.459.$
Later we shall see that the parameter $\beta $ controls the sign and magnitude of the
electric field in the outer region. The field cannot be induced,
when the diffusion coefficients of ions are equal and $\beta $
vanishes.

The no-slip boundary condition implies that at the surface fluid and particle velocities are equal, $\mathbf{V}=U\mathbf{e}_{x}$,
where $U$ is the dimensionless propulsion velocity along the $x-$axis scaled similarly to $\mathbf{V}$.
Finally, we set the value of the (constant) surface potential $\phi _{s}$
and naturally require $\Phi =\phi _{s}$, when $r=1$.

The boundary conditions at $r\rightarrow \infty $ are the conditions on the
concentration and potential~\cite{prieve1984motion},%
\begin{eqnarray}
C^{\pm } &=&1+\varepsilon r\cos \theta ,\   \label{bc_f_inf} \\
\Phi &=&-\beta \varepsilon r\cos \theta ,  \label{bc_fi}
\end{eqnarray}%
and the zero fluid velocity. The condition on potential (\ref{bc_fi})
ensures a zero current density far from the sphere.

\subsection{Outer solution}

Since $\lambda \ll 1,$ the leading-order solution to Eq. (\ref{pois1}) in
the outer region is $C^{+}=C^{-}=C,$ i.e. the electroneutrality holds to $%
O\left( \lambda ^{2}\right) .$ Equations (\ref{NPH},\ref{NP1}) then become
\begin{eqnarray}
\Delta C+\nabla \cdot \left( C\nabla \Phi \right) &=&0,  \label{NP2} \\
\Delta C-\nabla \cdot \left( C\nabla \Phi \right) &=&0.  \label{NP3}
\end{eqnarray}%
Summing up and subtracting Eqs. (\ref{NP2}) and (\ref{NP3}) along with the
boundary conditions (\ref{bc_11}) we obtain \cite%
{nourhani2015,ibrahim2017,asmolov2022self}:
\begin{eqnarray}
\Delta C &=&0,  \label{dif} \\
\nabla \cdot \left( C\nabla \Phi \right) &=&0,  \label{pot2}
\end{eqnarray}%
and for $r=1$:
\begin{gather}
\partial _{r}C=-\mathrm{Da},  \label{bc1} \\
C\partial _{r}\Phi =\beta \mathrm{Da}.  \label{bc_p}
\end{gather}

Solution to Eq. (\ref{dif}) is sought in the form $C_{0}\left( r\right)
+\varepsilon C_{1}\left( r\right) \cos \theta ,$ where the first term is a
spherically symmetric field induced by the surface flux and the second one
represents a small disturbance caused by external gradients. Both terms
satisfy the Laplace equation,%
\begin{equation}
\Delta C_{0}=0,\quad \Delta \left( C_{1}\cos \theta \right) =0.  \label{phi1}
\end{equation}

It follows from Eqs.~(\ref{bc_f_inf}) and (\ref{bc1}) that the boundary
conditions to (\ref{phi1}) can be formulated as
\begin{equation}
r=1:\ \ \partial _{r}C_{0}=-\mathrm{Da},\quad \partial _{r}C_{1}=0,
\label{y10}
\end{equation}%
\begin{equation}
r\rightarrow \infty :\ C_{0}=1,\quad C_{1}=r.  \label{bci2}
\end{equation}%
Solutions to Eqs. (\ref{phi1})-(\ref{bci2}) can then be readily obtained:
\begin{equation}
C_{0}=1+\frac{\mathrm{Da}}{r},  \label{c_sol}
\end{equation}%
\begin{equation}
C_{1}=r+\frac{1}{2r^{2}}.  \label{c_sol1}
\end{equation}%
These equations indicate that the concentration disturbance $C_{1}$ is the
same as for a non-catalytic particle, but the main-order concentration $%
C_{0} $ differs.

The electrostatic potential followed from Eq. (\ref{pot2}) can be expressed
in terms of $C$. Note that the boundary conditions both at the particle
surface (\ref{bc_p}) and at infinity (\ref{bc_fi}) are equivalent to a
condition of vanishing electric current. Whence the current is zero over the
whole outer region. Thus, the solution to Eq. (\ref{pot2}) obtained for
self-diffusiophoresis of Janus particles \cite%
{asmolov2022self,asmolov2022MDPI,asmolov2023PhF},
\begin{equation}
\Phi =-\beta \ln \left( C\right) ,  \label{fi_sol}
\end{equation}%
remains valid for our case of an imposed external concentration gradient
too. Equation (\ref{fi_sol}) implies that the ion fluxes induce a finite
potential difference in the outer region. Therefore, in contrast to a
non-catalytic particle, the concentration and potential at the outer edge of
the inner region, $C_{s}$ and $\Phi _{s},$ may differ considerably from the
bulk values, $C=1$ and $\Phi =0$,%
\begin{equation}
C_{s}=C_{0}\left( 1\right) +\varepsilon C_{1}\left( 1\right)
=C_{s0}+\varepsilon \frac{3}{2}\cos \theta ,  \label{cs0}
\end{equation}%
\begin{equation}
\Phi _{s}=\Phi _{0}\left( 1\right) +\varepsilon \Phi _{1}\left( 1\right)
=\Phi _{s0}-\beta \varepsilon \frac{3}{2C_{s0}}\cos \theta ,  \label{cs}
\end{equation}%
\begin{equation*}
C_{s0}=1+\mathrm{Da,\quad }\Phi _{s0}=-\beta \ln \left( C_{s0}\right) .
\end{equation*}%
We remark and stress that the surface concentration is  $c_{\infty
}+\frac{ja}{D},\ $ so it may remain finite even in highly dilute
solutions, provided $\mathrm{Da}\gg 1.$ The screening length near the particle can be evaluated as $\lambda _{D}\left( 1+\mathrm{Da}\right) ^{-1/2}$ and is smaller than the Debye length of the bulk solution. This suggests that the condition of a thin EDL should rather be $\lambda \left( 1+\mathrm{Da}\right) ^{-1/2}\ll 1$
and this could obey even in very dilute solutions as well.

The outer charge distribution which generates an electric field can be
calculated using (\ref{pois1}):%
\begin{equation}
\varrho =C_{0}^{+}-C_{0}^{-}=2\lambda ^{2}\beta \Delta \ln \left( C_{0}\right)
=-2\lambda ^{2}\beta \frac{\mathrm{Da}^{2}}{r^{2}\left( r+\mathrm{Da}\right)
^{2}}.  \label{char0}
\end{equation}%
Clearly, the outer charge can only arises if anions and cations are of  different
diffusivities (finite $\beta$ and Da). It
is small compared to the charge inside the EDL, being of the order of $\lambda ^{2}%
\mathrm{Da}^{2}\ll 1,$ but the size of the charged cloud is of the order of the
particle size and significantly exceeds the dimension of the inner (EDL)
region. This could have important consequences for diffusiophoretic
propulsion as we demonstrate below.

\subsection{Inner solution}

The inner solution can be constructed by using a stretched
coordinate $\rho =(r-1)/\lambda $. The dimensionless potential and
concentrations in the inner region are sought in the form $\phi =\Phi
_{s}+\varphi  $ and $C^{\pm }=C_{s}\xi ^{\pm },$ respectively. The outer limits of the inner
solution are then obtained by matching with the inner limits of the outer
solution, Eqs.\eqref{cs0} and \eqref{cs}, %
\begin{gather}
\rho \rightarrow \infty :\quad \xi ^{\pm }=1,  \label{ou_l} \\
\qquad \qquad \varphi =0 .
\notag
\end{gather}

The boundary condition for ion fluxes (\ref{bc_11}) written in
terms of $\rho ,$ reads %
\begin{equation*}
\rho =0:\quad \lambda ^{-1}\left( C_{s}\partial _{\rho }\xi ^{\pm }\pm
C_{s}\xi ^{\pm }\partial _{\rho }\varphi \right) =-\mathrm{Da}\frac{1\mp
\beta }{2},
\end{equation*}%
so we obtain $\partial _{\rho }\xi ^{\pm }\left( 0\right) \pm \xi
^{\pm }\partial _{\rho }\varphi \left( 0\right) \sim \lambda \ll 1.$ Therefore, in the inner region the fluxes can safely be neglected, and we
can then conclude that the ion concentration fields satisfy the Boltzmann
distribution,%
\begin{equation}
\xi ^{\pm }=\exp \left( \mp \varphi \right) ,  \label{Bol}
\end{equation}%
and, consequently, that the potential obeys the Poisson-Boltzmann
equation %
\begin{equation}
\partial _{\rho \rho }\varphi =C_{s}\sinh \varphi .  \label{PB}
\end{equation}%
At the surface, $\rho =0$, the boundary condition to (\ref%
{PB}) reads %
\begin{equation}
\varphi (0)=\phi _{s}-\Phi _{s}\equiv \psi .  \label{phi_s}
\end{equation}%
Here $\psi $ is the potential jump in the inner region.
Therefore, the inner-problem equations and boundary conditions are similar
to those for inert particle, but the bulk values, $C=1$ and $\Phi
=0,$ should be replaced by $C_{s0}$ and $\Phi _{s0}.$ As a
side note, $\psi $ is similar, but not fully identical to a so-called
zeta-potential, which for hydrophilic surfaces represents itself a potential
jump in a whole EDL and is equal to $\phi _{s}$.

\subsection{Particle velocity}

The velocity of a freely moving in the $x-$direction particle can be
determined by using the reciprocal theorem \cite{teubner1982}
\begin{equation}
U=-\frac{1}{6\pi }\int_{V_{f}}\mathbf{f}\cdot \left( \mathbf{V}_{St}-\mathbf{%
e}_{x}\right) dV.  \label{rec}
\end{equation}%
The integral is evaluated over the whole fluid volume $V_{f}$ and $\mathbf{V}%
_{St}\left( \mathbf{r}\right) $ represents the velocity field for the
particle of the same radius that translates with the unit velocity $\mathbf{e%
}_{x}$ in a stagnant fluid along the $x-$axis (Stokes solution):
\begin{equation}
\mathbf{V}_{St}=\left( \frac{3}{2r}-\frac{1}{2r^{3}}\right) \cos \theta
\mathbf{e}_{r}-\left( \frac{3}{4r}+\frac{1}{4r^{3}}\right) \sin \theta
\mathbf{e}_{\theta },  \label{v_st}
\end{equation}%
where $\mathbf{e}_{r}$ and $\mathbf{e}_{\theta }$ are the unit vectors. The
leading-order solution (\ref{c_sol}) does not induce any velocity because of
its symmetry. The first-order particle velocity can be presented as a
superimposition of the contributions of the inner and the outer regions \cite%
{asmolov2022self},
\begin{equation}
U=\varepsilon \left( U_{i}+U_{o}\right) .  \label{v_full}
\end{equation}%
Both regions, $r-1\sim \lambda $ and $r\sim 1$, contribute to the particle
velocity when $\mathrm{Da\sim 1}$. However, at large \textrm{Da} the
contribution from an outer region becomes dominant. Indeed, we can deduce
from Eq.~(\ref{char0}) that the charge density $\rho =O\left( \lambda ^{2}%
\mathrm{Da}^{-2}\right) $ when $r\sim \mathrm{Da}\gg 1,$ while the region
volume grows like $\mathrm{Da}^{3}.$ Thus we might argue that the particle
speed should grow linearly with $\mathrm{Da}$. Later we shall see that this
is indeed so.

The first term in Eq.~(\ref{v_full}) is associated with the slip velocity. For a diffusiophoretic migration of passive non-catalytic particles ($\Da=0$)
it is given by the known formula~\cite%
{prieve1984motion,anderson1989colloid}
\begin{equation}
U_{i0}=\beta \phi _{s} +4\ln \left[ \cosh \left( \frac{%
\phi _{s} }{4}\right) \right].  \label{ui0}
\end{equation}%
Here the first and the second terms are associated
with electro- and diffusio-osmotic flows, respectively. The electroosmotic contribution vanishes for equal ion diffusivities ($\beta =0$).
For non-zero Da, similarly to Eq. \eqref{ui0} one can write 
\begin{equation}
U_{i}=\frac{1}{C_{s0}}\left\{ \beta \psi +4\ln \left[ \cosh \left( \frac{%
\psi }{4}\right) \right] \right\},  \label{ui}
\end{equation}%
where we take into account that the concentration and potential at the outer edge of the inner region (EDL), $C_{s0}$ and $\Phi _{s0}$, differ from their bulk values.
This expression coincides
with Eq. \eqref{ui0} when $\mathrm{Da}=0,$  and the
correction to $U_{i}$ is $O\left( \mathrm{Da}\right) $ as $\mathrm{Da}\ll 1$.
In the opposite limit of $\mathrm{Da}\gg 1$, we obtain $C_{s0}\simeq
\mathrm{Da}$ and $\psi \simeq \beta \ln \left( \mathrm{Da}\right) $. It
follows then from (\ref{ui}) that the velocity is decaying with $\mathrm{Da:}
$%
\begin{equation}
U_{i}\simeq \left( \beta ^{2}+\left\vert \beta \right\vert \right) \frac{\ln
\left( \mathrm{Da}\right) }{\mathrm{Da}}\ll 1\text{\quad as\quad }\mathrm{Da}%
\gg 1.  \label{uias}
\end{equation}

The second term in \eqref{v_full} that is associated with an outer region
can be rewritten using Eqs. (\ref{fi_sol}) and (\ref{rec}) as \cite%
{asmolov2022self}
\begin{equation}
U_{o}=-\frac{\beta ^{2}}{6\pi }\int_{V_{f}}\frac{\left( \mathbf{\nabla }%
C\right) ^{2}\mathbf{\nabla }C\cdot \left( \mathbf{V}_{St}-\mathbf{e}%
_{x}\right) }{C^{3}}dV,  \label{uo}
\end{equation}%
where the integral is taken over the outer region. By expanding Eq. (\ref{uo}%
) in $\varepsilon $ one can obtain%
\begin{equation}
U_{o}=-\frac{1}{3}\int_{1}^{\infty }\int_{0}^{\pi }\mathbf{f}_{1}\cdot
\left( \mathbf{V}_{St}-\mathbf{e}_{x}\right) r^{2}\sin \theta d\theta dr,
\label{u1o}
\end{equation}%
where
\begin{eqnarray}
\mathbf{f}_{1} &=&\frac{\beta ^{2}}{C_{0}^{3}}\left[ \left\{ 2\mathbf{\nabla
}C_{0}\cdot \mathbf{\nabla }\left( C_{1}\cos \theta \right) -\frac{%
3C_{1}\cos \theta \left( \mathbf{\nabla }C_{0}\right) ^{2}}{C_{0}}\right\}
\mathbf{\nabla }C_{0}\right.  \notag \\
&&\left. \frac{{}}{{}}+\left( \mathbf{\nabla }C_{0}\right) ^{2}\mathbf{%
\nabla }\left( C_{1}\cos \theta \right) \right] .  \label{f1b}
\end{eqnarray}

The velocity $U_{o}$ is $O\left( \mathrm{Da}^{2}\right) $ when $\mathrm{Da}%
\ll 1$ since it is quadratic in concentration gradient $\mathbf{\nabla }%
C_{0} $ which is proportional to $\mathrm{Da.}$ At large $\mathrm{Da,}$ we
can obtain from Eq.~(\ref{c_sol}) that $\mathbf{f}_{1}=O\left( \mathrm{Da}%
^{-1}\right) $ for $r\sim 1.$ However, large distances $r\sim \mathrm{Da}\gg
1$ provide much greater contribution to the volume integral (\ref{u1o}). The
integrand of (\ref{u1o}) can be evaluated at large $r$ as
\begin{equation}
-\frac{\beta ^{2}\mathrm{Da}^{2}r}{\left( r+\mathrm{Da}\right) ^{3}}\left[
3\cos ^{2}\theta \left( 1+\frac{\mathrm{Da}}{r+\mathrm{Da}}\right) +\sin
^{2}\theta \right]\sin \theta .  \label{f1_a}
\end{equation}%
As a result, one can calculate (\ref{u1o}) explicitly:%
\begin{equation}
U_{o}\simeq \frac{2\beta ^{2}\mathrm{Da}}{3}\quad \text{as}\quad \mathrm{Da}%
\gg 1,  \label{uo_as}
\end{equation}
i.e. the velocity $U_{o}$ is linear in $\mathrm{Da}.$

\section{Results and discussion}

\label{s3}

In this section we calculate numerically the diffusiophoretic velocity given
by Eqs. (\ref{v_full}), (\ref{ui}) and (\ref{u1o}) as a function of main
dimensionless parameters of our problem.

\begin{figure}[th]
\centering
\includegraphics[width=1\columnwidth ]{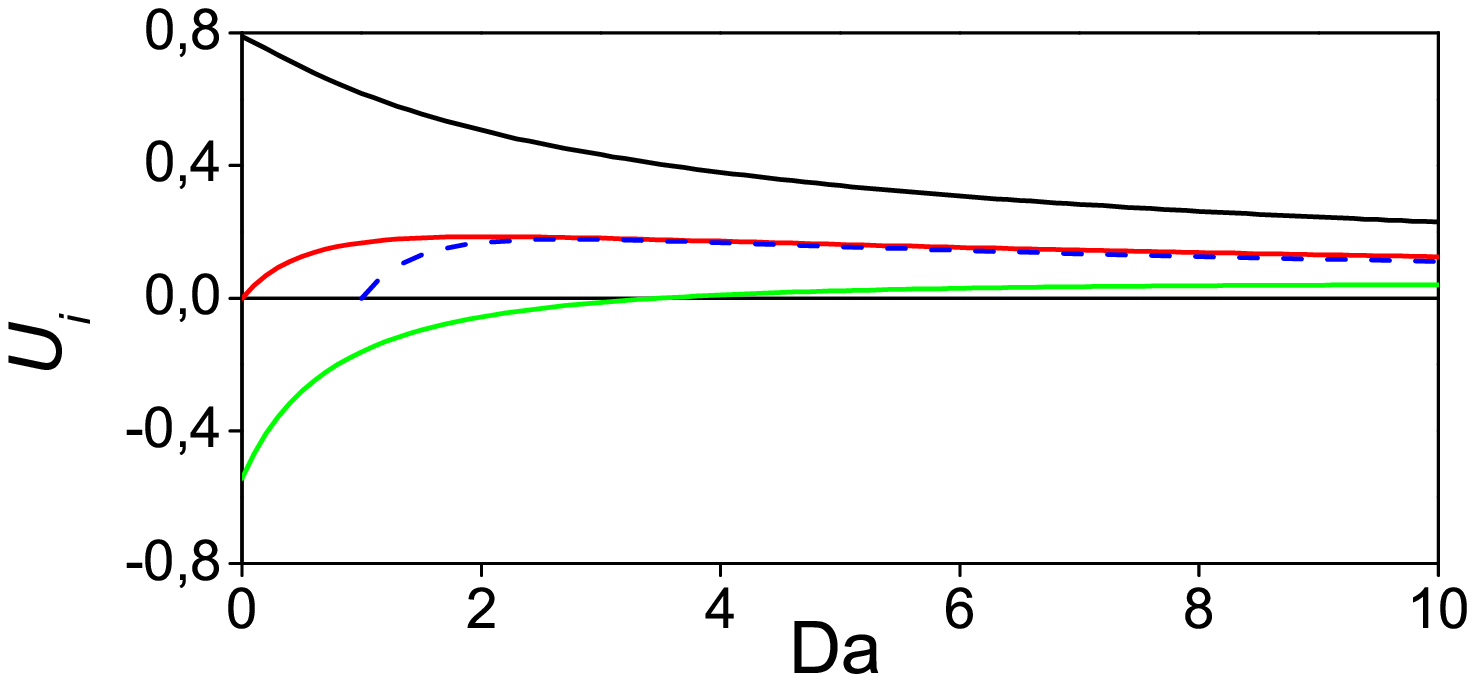} 
\vspace{-2.4cm}
\caption{Contribution of the inner region to the particle velocity vs. Da
calculated using $\protect\phi _{s}=1$, $0$, $-1$ (from top to bottom) and $%
\protect\beta =2/3$ (solid curves). Dashed curve corresponds to asymptotic
solution \eqref{uias}. }
\label{fig:vi}
\end{figure}

We begin by studying the contribution of the inner region to the velocity
given by Eq. (\ref{ui}). This contribution depends on three parameters.
Namely, on the two restricted $\left\vert \beta \right\vert <1,$ $\mathrm{Da}%
\geq 0,$ and on the surface potential $\phi _{s}$, which, in principle, may
be any. Since the velocity $U_{i}$ is invariant to the simultaneous change
in signs of $\beta $ and $\phi _{s}$~\cite{asmolov2022self}:%
\begin{equation}
U_{i}\left( -\beta ,\mathrm{Da},-\phi _{s}\right) =U_{i}\left( \beta ,%
\mathrm{Da},\phi _{s}\right) ,  \label{sign}
\end{equation}%
we calculate $U_{i}$ using only positive $\beta $, but set $\phi _{s}$ of
both signs. The results are plotted in Fig.~\ref{fig:vi} as a function of
Damk\"{o}hler number. The sign of $U_{i}$ coincides with that of a potential
drop $\psi $ across the EDL that is given by \eqref{phi_s}. For a
non-catalytic particle $\mathrm{Da}=0$, so is $\Phi _{s}$. Thus $\psi =\phi
_{s}$, and the velocity is of the same sign as the surface potential. This
means that a positively charged particle migrates towards a more
concentrated solution, but a negatively charged - to a diluter one.
Naturally, the neutral particle of $\phi _{s}=0$ remains immobile. On
increasing $\mathrm{Da}$, however, the classical behavior violates. The
point is that $\Phi _{s}$ becomes negative (see Eq.~(\ref{cs})) and,
therefore, increases $\psi $. As a result, the velocity of a neutral
particle becomes positive even for small Damk\"{o}hler number, and the
negatively charged particle reverses its migration direction to a positive
at some finite (moderate) $\mathrm{Da}$. If $\mathrm{Da}$ is large,
(positive) velocities of particles decay weakly with $\mathrm{Da}$ according
to Eq.~(\ref{uias}).

\begin{figure}[th]
\centering
\includegraphics[width=1\columnwidth ]{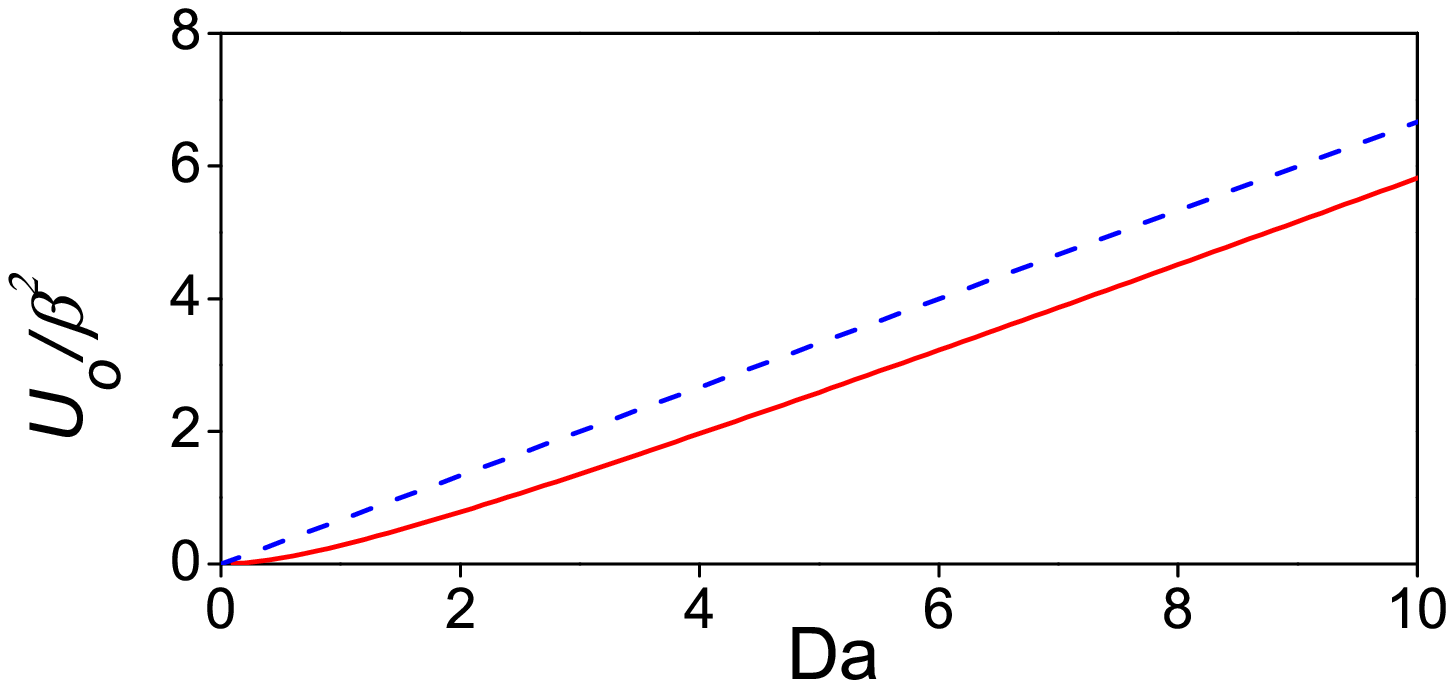} \vspace{-2.0cm}
\caption{Contribution of the outer region to the particle velocity vs. Da
(solid curve). Dashed line shows calculations from Eq.~\eqref{uo_as}.}
\label{fig:vo}
\end{figure}

Let us now turn to the contribution of the outer region to the particle
speed. Recall that $U_{o}$ is given by Eq.~(\ref{u1o}) and is quadratic in $%
\beta ,$ but the integral in this equation depends on $\mathrm{Da}$ solely.
To perform the integration, we first define a new variable $s=r^{-1}$. The
integral is then computed on uniform grids in $\theta $\ and $s$ using $%
N_{\theta }=N_{s}=200$. The results presented in Fig.~\ref{fig:vo} show that
$U_{o}$ demonstrates a quadratic growth for small $\mathrm{Da}$ values, but
augments linearly for larger Damk\"{o}hler numbers. Also included in Fig.~%
\ref{fig:vo} is the calculation from asymptotic Eq.~(\ref{uo_as}). It can be
seen that it slightly overestimates $U_{o}$. Nevertheless, the asymptotic
calculation demonstrates the magnitude of the outer contribution and correct
scaling with $\mathrm{Da}$, when the latter becomes large.

\begin{figure}[th]
\vspace{-0.55cm}
\centering
\begin{minipage}[h]{1\linewidth}
\includegraphics[width=1.6\columnwidth ]{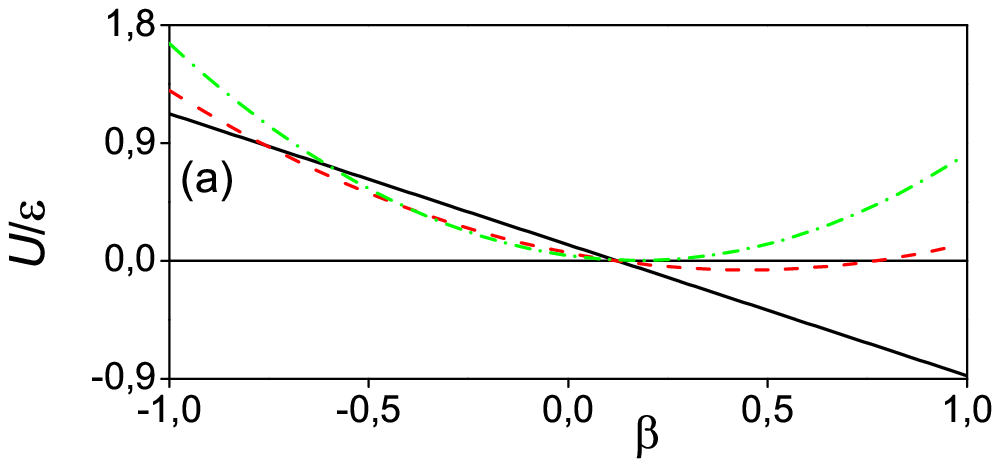}
\end{minipage}
\vfill
\begin{minipage}[h]{1\linewidth}
\vspace{-5.7cm}
\includegraphics[width=1.6\columnwidth ]{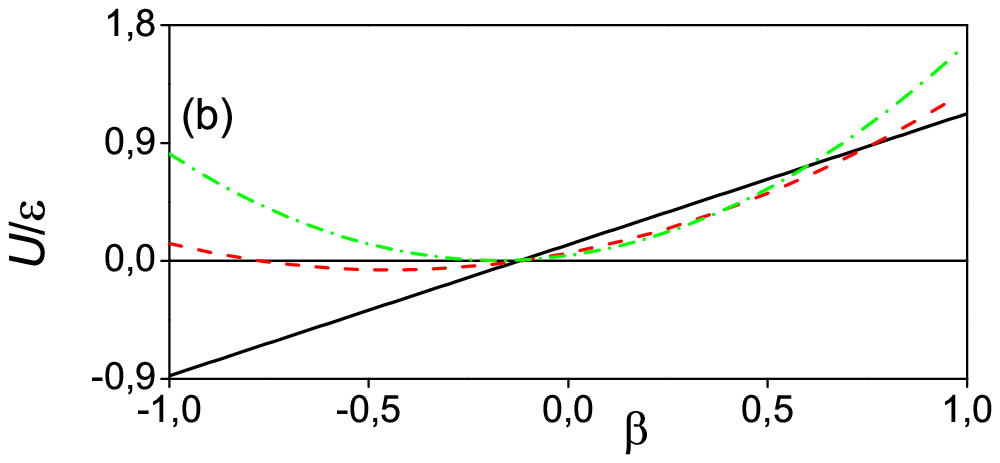}
\end{minipage}
\vspace{-5.7cm}
\caption{Total particle velocity vs. $\protect\beta $ for (a) $\protect\phi _{s}=-1$ and (b) $\protect\phi _{s}=1$.
Solid, dashed and dashed-dotted curves correspond to  $\Da =0$, $1 $, $2$. }
\label{fig:bet}
\end{figure}

\begin{figure}[th]
\centering
\begin{minipage}[h]{1\linewidth}
\includegraphics[width=1\columnwidth ]{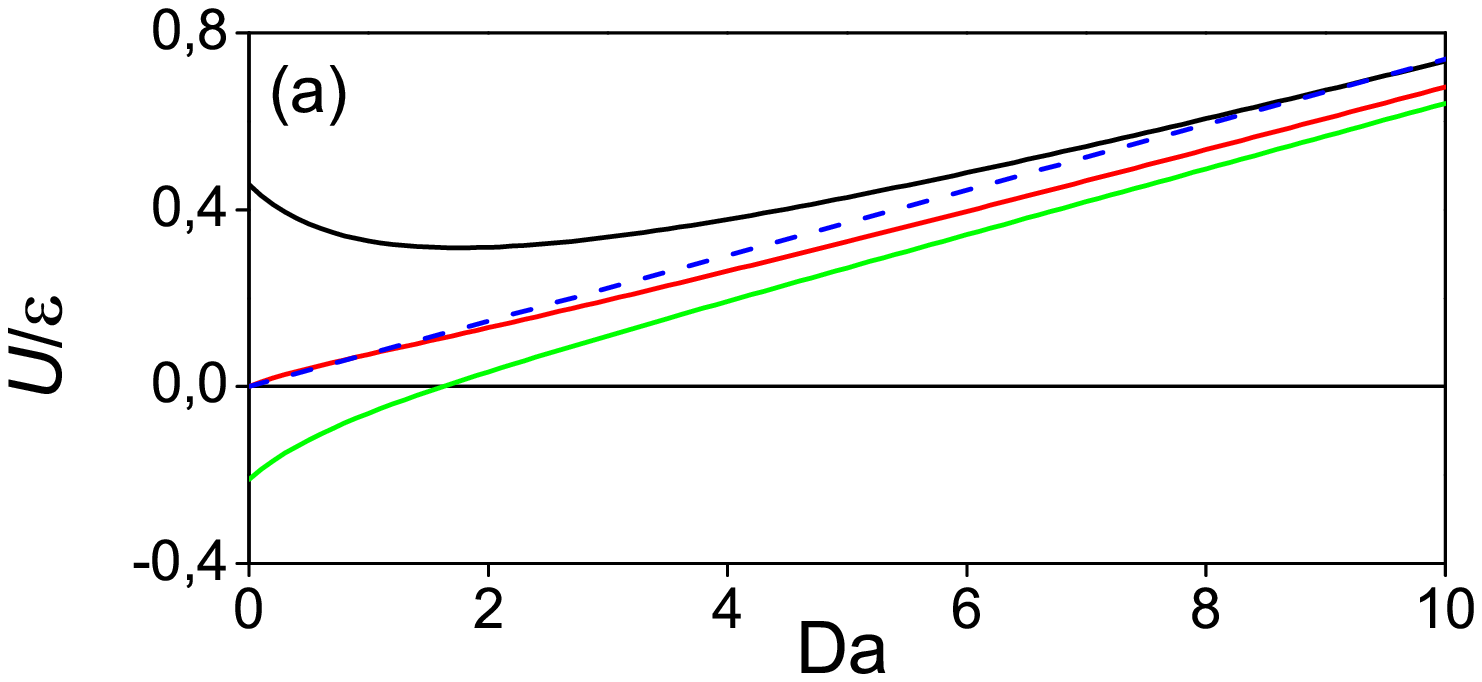}
\end{minipage}
\vfill
\begin{minipage}[h]{1\linewidth}
\vspace{-2.4cm}
\includegraphics[width=1\columnwidth ]{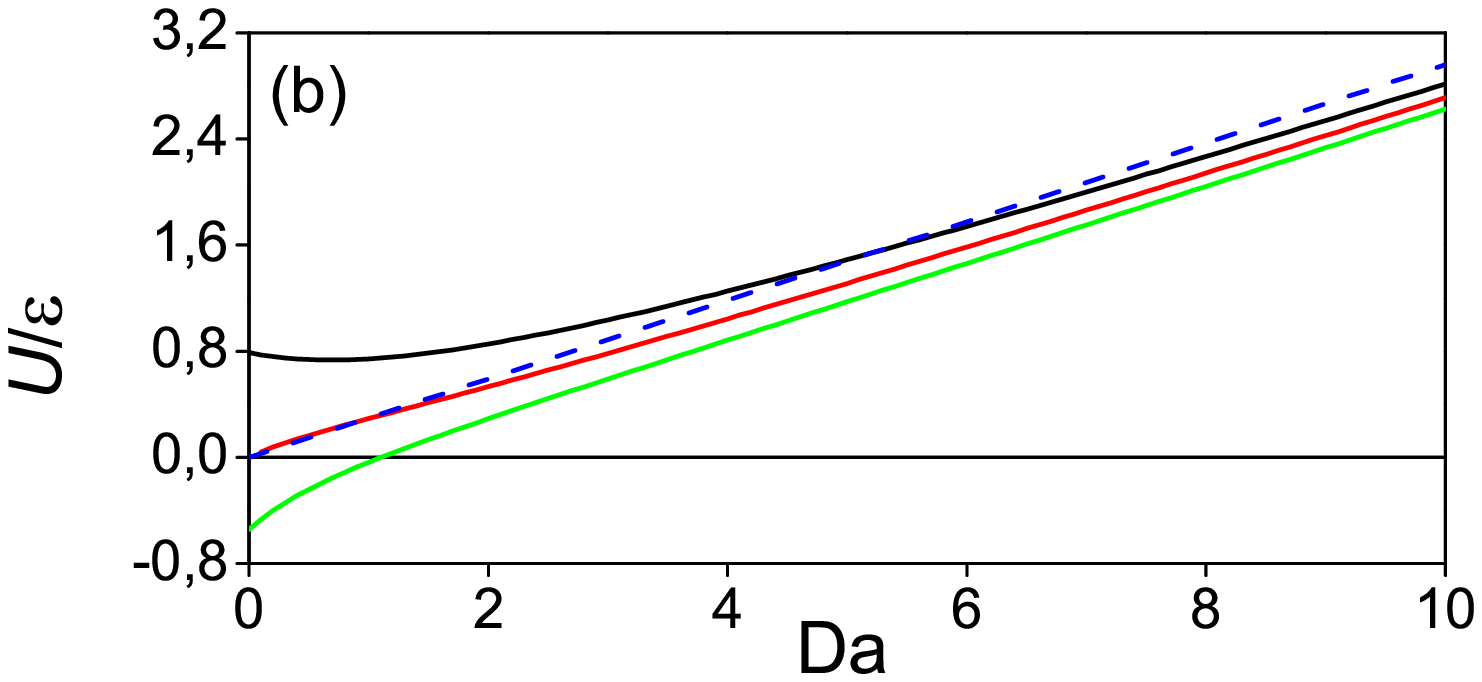}
\end{minipage}
\vspace{-2.0cm}
\caption{Total particle velocity vs. Da for (a) $\protect\beta =1/3$ and (b) $%
\protect\beta =2/3$. Solid lines are calculated for $\protect\phi _{s}=1$, $%
0 $, $-1$ (from top to bottom). Dashed line corresponds to asymptotic
solution~\eqref{uo_as}.}
\label{fig:net}
\end{figure}

Finally, we compute the diffusiophoretic velocity of the particle $U$ given
by Eq.~\eqref{v_full}.
Figure \ref{fig:bet} shows the total particle velocities as a function of $%
\beta $ calculated using several Da and $\phi _{s}.$ It can be seen that (\ref{sign}) remains valid for the total velocity $U$ too.
For $\beta =0$ (equal ion diffusivities), both electroosmotic contribution
(the first term in (\ref{ui})) and the contribution of the outer region $%
U_{o}$ vanish as they are proportional to $\beta $ and $\beta ^{2},$
respectively. As a result, only second term in (\ref{ui}) contributes to the
velocity. It is positive, but relatively small. For $%
\mathrm{Da}=0$ (Eq. \eqref{ui0} for non-catalytic particles, solid lines in Fig. \ref{fig:bet}),
$U$ is given by (\ref{ui0}) and is linear in $\beta $ due to the
electroosmotic term. The sign of $U$ and, consequently, the direction of motion are
controlled by $\beta $ and $\phi _{s}:$ the particle translates along $x-$axis
when they are of the same signs, and in the opposite direction if their sign are different. For a moderate Damk\"{o}hler number ($\mathrm{Da}=1
$, dashed lines in Fig. \ref{fig:bet}), finite positive velocities are
observed only if $\beta$ and $\phi _{s}$ are both positive or negative, otherwise their magnitudes are very small. For greater $\mathrm{%
Da}$ ($\mathrm{Da}=2$, dashed-dotted lines), the velocity is always positive
due to the growing contribution of the outer region.

The calculations made using $\beta =1/3$ and $2/3$
and the same values of $\phi _{s}$ as in Fig.~\ref{fig:vi} are illustrated  in Fig.~\ref{fig:net}. Theoretical
curves given by ~\eqref{uo_as} are included. Our results also show that one can
define two different regimes depending on the value of the Damk\"{o}hler
number. When $\mathrm{Da}$ is sufficiently small, the velocity is still to a
large extent controlled by $\phi _{s}$ like it is for non-catalytic
particles. This is not surprising since in this regime $U$ is mainly
determined by the slip velocity at the edge of the EDL (i.e. of the inner
region), but the impact of the outer volume charge is quite small. Indeed, $%
O\left( \mathrm{Da}^{2}\right) $, when $\mathrm{Da}$ is small, as discussed
above. However, when $\mathrm{Da}$ becomes large enough, the velocities of
particles are positive, even if they are neutral or negatively charged. In
other words, at large $\mathrm{Da}$ any catalytic particle migrates in the
direction of higher concentration, which is consistent with experimental
observations~\cite{dey2015micromotors}. We also remark that all curves in
Fig.~\ref{fig:net} converge to a single one, and are rather well fitted by
Eq.~(\ref{uo_as}). Note that in this mode the contribution of the outer
region $U_{o}$ dominates, so that this convergence would become more pronounced for larger $\beta,$ since $U_{o}\propto
\beta ^{2}$ while $U_{i}\propto \beta .$ These results suggest
that the particle velocity is controlled mostly by the Damk\"{o}hler number,
but not by the surface potential as it would be in the case of inert
non-catalytic particles or at a very small Da. Thus, at $\mathrm{Da}\gg 1$,
the dimensionless migration velocity can be estimated as%
\begin{equation}
U\sim \varepsilon \mathrm{Da}=\frac{Ja}{Dc_{\infty }^{2}}\frac{dc_{\infty }}{%
dx}\equiv M\frac{dc_{\infty }}{dx},  \label{eq:Ufinal}
\end{equation}%
where $M$ is a diffusiophoretic mobility of a particle. It follows from Eq.~%
\eqref{eq:Ufinal} that in the large $\mathrm{Da}$ regime the mobility of
passive catalytic particles is proportional to $c_{\infty }^{-2}$, which is
different from $M\sim c_{\infty }^{-1}$ derived for non-catalytic particles~%
\cite{prieve1984motion,anderson1989colloid}. The consequences of this
scaling can be huge. We recall that large $\mathrm{Da}$ can be achieved at
low $c_{\infty }$ (see Eq. (\ref{Da_def})). Our scaling thus predicts that
in very dilute solutions catalytic particles would migrate much faster than
inert ones.

\section{Conclusion}

\label{s4}

We proposed a theory of a diffusiophoresis of a catalytic particle with an
uniform ion release. The theory is valid in the limit of a thin
electrostatic diffuse layer and employs the method of matched asymptotic
expansions. Our model provides considerable insight into two different
regimes of a diffusiophoretic migration of passive catalytic particles. In
the regime, which is observed at the low Damk\"{o}hler number, the surface
potential $\phi _{s}$ controls to a large extent the velocity of the
particle and its direction. In the other regime, where Da is high, particles
migrate towards a high concentration region with a velocity, which scales
with Da and depends neither on the magnitude of a surface potential, nor on
its sign. In this regime the diffusiophoretic mobility becomes proportional
to $c_{\infty }^{-2},$ which differs from the scaling expression $c_{\infty
}^{-1}$ obtained earlier for non-catalytic particles and allows one to argue
that in dilute solutions catalytic particles would migrate much faster than
inert.

Our strategy can be extended to more complex situations, such as, for
example an electrophoresis of passive catalytic particles, under an applied
electric field. So far, only an electrophoretic migration of inert passive
particles has been investigated, and it would be very timely to understand
what kind of changes are expected if passive particles are catalytic.
Another fruitful direction would be to consider a self-propulsion of active
Janus particles in the external concentration fields.

\begin{acknowledgments}

This work was supported by the Ministry of Science and Higher Education of the Russian Federation.
\end{acknowledgments}

\section*{DATA AVAILABILITY}

The data that support the findings of this study are available within the
article.

\section*{AUTHOR DECLARATIONS}

The authors have no conflicts to disclose.

\bibliographystyle{unsrt}
\bibliography{dph,current}

\end{document}